\documentclass[10pt]{article}
\usepackage{epsfig}
\usepackage{amssymb}
\usepackage{amsmath}

\usepackage{txfonts}
\usepackage{graphicx}
\usepackage{color}
\usepackage{colortab}
\usepackage{pstricks}
\usepackage{enumerate}
\usepackage[normalem]{ulem}

\usepackage{natbib}
\bibpunct{(}{)}{;}{a}{}{,} 
\definecolor{seda}{gray}{.83}
\newcommand*\seda{\color{seda}}

\begin{document}
\begin{center}
{{\bf\Large Resonant switch model of twin peak HF QPOs\\
applied to the source 4U~1636$-$53}}\\
\vspace{3ex}
 {\large Zden\v{e}k Stuchl\'{\i}k, Andrea Kotrlov\'a and Gabriel T\"{o}r\"{o}k}\\
\vspace{3ex}
{Institute of Physics, Faculty of Philosophy and Science, Silesian University in Opava, Bezru\v{c}ovo n\'{a}m. 13,
CZ-74601 Opava, Czech Republic\\
{\small e-mail: (zdenek.stuchlik, andrea.kotrlova)@fpf.slu.cz}}
\end{center}

\vspace{3ex}
\begin{center}
{ABSTRACT}
\end{center}
\noindent{\small Resonant Switch (RS) model of twin peak high-frequency quasi-periodic oscillations (HF~QPOs) assumes switch of twin oscillations at a~resonant point where frequencies of the~upper and lower oscillations $\nu_{\mathrm{U}}$ and $\nu_{\mathrm{L}}$ become commensurable and the~twin oscillations change from one pair of the~oscillating modes (corresponding to a~specific model of HF~QPOs) to some other pair due to non-linear resonant phenomena. The RS model is used to determine range of allowed values of spin $a$ and mass $M$ of the~neutron star located at the~atoll source 4U~1636$-$53 where two resonant points are observed at frequency ratios $\nu_{\mathrm{U}} : \nu_{\mathrm{L}} = 3\!:\!2$, 5\,:\,4. We consider the~standard specific models of the~twin oscillations based on the~orbital and epicyclic geodetical frequencies. The~resonant points are determined by the~energy switch effect exhibited by the~vanishing of the~amplitude difference of the~upper and lower oscillations. The~predicted ranges of the~neutron star parameters are strongly dependent on the~twin modes applied in the~RS~model. We demonstrate that for some of the~oscillatory modes used in the~RS~model the~predicted parameters of the~neutron star are unacceptable. Among acceptable RS~models the~most promising are those combining the~Relativistic Precession and the~Total Precession frequency relations or their modifications.}


\section{Introduction}

In the~Galactic Low Mass X-Ray Binaries (LMXB) containing neutron (quark) stars quasiperiodic oscillations (\emph{QPOs}) of X-ray brightness had been observed at {low}-(Hz) and {high}-(kHz) frequencies \citep[see, e.g.,][]{Kli:2000:ARASTRA:,Kli:2006:CompStelX-Ray:,Bar-Oli-Mil:2005:MONNR:}. Since the~high frequencies are close to the~orbital frequency of the~marginally stable circular orbit representing the~inner edge of Keplerian disks orbiting neutron stars, the~strong gravity effects must be relevant in explaining HF~QPOs.

The~HF~QPOs in neutron star systems are often demonstrated as two simultaneously observed pairs of peaks (twin peaks) in the~Fourier spectra~corresponding to oscillations at the~upper and lower frequencies $(\nu_{\mathrm{U}}, \nu_{\mathrm{L}})$. The~twin peaks at the~upper and lower frequencies $(\nu_{\mathrm{U}}, \nu_{\mathrm{L}})$ substantially change over time (in one observational sequence). Sometimes only one of the~frequencies is observed and evolved \citep[][]{Bel-etal:2007:MONNR:RossiXTE,Bou-Bar-Lin-Tor:2010:MNRAS:}. Most of the~twin HF~QPOs in the~so called atoll sources \citep[][]{Kli:2006:CompStelX-Ray:} have been detected at lower frequencies $600-800\,\mathrm{Hz}$ vs. upper frequencies $900-1200\,\mathrm{Hz}$,
demonstrating a~clustering of the~twin HF~QPOs frequency ratios around 3\,:\,2. This clustering indicates some analogy to the~black hole (BH) case where twin peaks with fixed pair of frequencies at the~ratio 3\,:\,2 are usually observed \citep[][]{Abr-etal:2003:ASTRA:,Bel-etal:2007:MONNR:RossiXTE,Tor-etal:2008:ACTA:DistrKhZ4U1636-53,Tor-Bak-Stu-Cec:2008:ACTA:TwPk4U1636-53,Bou-Bar-Lin-Tor:2010:MNRAS:}. Amplitudes of twin HF~QPOs in the~neutron star systems are usually stronger and their coherence times are higher than those of BH sources \citep[see, e.g.,][]{Rem-McCli:2006:ARASTRA:,Bar-Oli-Mil:2005:MONNR:}. It is probable that a~$3:2$ resonance plays a~significant role also in the~atoll sources containing neutron stars. However, this case is much more complicated, as the~frequency ratio, although concentrated around 3\,:\,2, falls in a much wider range than
in the BH systems \citep[][]{Ter-Abr-Klu:2005:ASTRA:QPOresmodel,Tor:2009:ASTRA:ReversQPOs,Bou-Bar-Lin-Tor:2010:MNRAS:,Mon-Zan:2012:MNRAS:}.

In fact, a~multi-peaked distribution in the~frequency ratios has been observed, i.e., more than one resonance could be realized if a~resonant mechanism is involved in generating the~neutron star HF~QPOs. For some atoll neutron star sources the~upper and lower HF~QPO frequencies can be traced along the~whole observed range, but the~probability to detect both QPOs simultaneously increases when the~frequency ratio is close to ratio of small natural numbers, namely 3\,:\,2, 4\,:\,3, 5\,:\,4 -- this has been observed in the~case of six atoll sources: 4U~1636$-$53, 4U~1608$-$52, 4U~0614$-$09, 4U~1728$-$34, 4U~1820$-$30, 4U~1735$-$44 \citep[][]{Tor:2009:ASTRA:ReversQPOs,Bou-Bar-Lin-Tor:2010:MNRAS:}. The~analysis of root-mean-squared-amplitude evolution in the~group of six atoll sources shows that the~upper and lower HF~QPO amplitudes equal each other and alter their dominance while passing rational frequency ratios corresponding to the~datapoints clustering \citep{Tor:2009:ASTRA:ReversQPOs}. Such an~``energy switch effect'' can be well explained in a~framework of non-linear resonant orbital models as shown in \citet{Hor-etal:2009:ASTRA:IntResQPOs}. Moreover, the~analysis of the~twin peak HF~QPO amplitudes in the~atoll sources (4U~1636$-$53, 4U~1608$-$52, 4U~1820$-$30, and 4U~1735$-$44) indicates a~cut-off at resonant radii corresponding to the~frequency ratios 5\,:\,4 and 4\,:\,3 respectively implying a~possibility that the~accretion disk inner edge is located at the~innermost resonant radius  rather than at the~innermost stable circular geodesic (ISCO) \citep{Stu-Kot-Tor:2011:ASTRA:ResRadKep}.

The~evolution of the~lower and upper twin HF~QPOs frequencies in the~atoll sources suggests (a~very rough) agreement of the~data distribution with so called hot spot models, especially with the~relativistic precession (RP) model prescribing the~evolution of the~upper frequency by $\nu_{\mathrm{U}}=\nu_\mathrm{K}$ and the~lower frequency by $\nu_{\mathrm{L}}=\nu_\mathrm{K} - \nu_r$ \citep{Ste-Vie:1999:PHYRL:,Ste-Vie:1998:ASTRJ2L:}. In rough agreement with the~data are also other models based on the~assumption of the~oscillatory motion of hot spots, or accretion disk oscillations, with oscillatory frequencies given by the~geodetical orbital and epicyclic motion: for example it holds for the~modified RP1 model with $\nu_{\mathrm{U}}=\nu_{\theta}$ and $\nu_{\mathrm{L}}=\nu_{\mathrm{K}} - \nu_r$ \citep{Bur:2005:RAGtime6and7:CrossRef}, the~total precession (TP) model with $\nu_{\mathrm{U}}=\nu_{\theta}$ (or $\nu_{\mathrm{U}}=\nu_\mathrm{K}$) and $\nu_{\mathrm{L}}=\nu_{\theta} - \nu_r$ \citep{Stu-Tor-Bak:2007:arXiv:}, the~tidal disruption (TD) model with $\nu_{\mathrm{U}}=\nu_\mathrm{K} + \nu_r$ and $\nu_{\mathrm{L}}=\nu_\mathrm{K}$ \citep{Cad-etal:2008:AA:,Kos-etal:2009:tidal:}, or the~warped disk oscillations (WD) model with $\nu_{\mathrm{U}}=2\nu_\mathrm{K} - \nu_r$ and $\nu_{\mathrm{L}}=2\left(\nu_{\mathrm{K}} - \nu_r\right)$ \citep{Kat:2004:PUBASJ:QPOsmodel,Kato:2008:b:PASJ:}. In all of these models the~frequency difference $\nu_{\mathrm{U}} - \nu_{\mathrm{L}}$ decreases with increase of the~magnitude of the~lower and upper frequencies in accord with trends given by the~observational data \citep{Bel-etal:2007:MONNR:RossiXTE,Bou-Bar-Lin-Tor:2010:MNRAS:}. This qualitative property of the~observational data excludes the~simple model of epicyclic oscillations with $\nu_{\mathrm{U}} = \nu_{\theta}$ and $\nu_{\mathrm{L}} = \nu_r$ \citep{Urb-etal:2010:ASTRA:DiscOscNS32} that works quite well in the~case of HF~QPOs in LMXB containing black holes \citep{Ter-Abr-Klu:2005:ASTRA:QPOresmodel}, or requires substantial modification of the model of epicyclic oscillations \citep{Abr-Klu-Yu:2011:AcA:}.

The~$\nu_\mathrm{U} / \nu_\mathrm{L}$ frequency relations, given by a variety of the~relevant frequency-relation models mentioned above, can be fitted to the~observational data for the~atoll sources containing neutron (quark) stars, e.g., data determined for the~atoll source 4U~1636$-$53 \citep{Bar-Oli-Mil:2005:MONNR:,Tor-etal:2008:ACTA:DistrKhZ4U1636-53,
Tor-Bak-Stu-Cec:2008:ACTA:TwPk4U1636-53}. The~parameters of the~neutron (quark) star spacetime can be then determined due to the~fitting of the~observational data by the frequency-relation models. The~rotating neutron stars are described quite well by the~Hartle--Thorne geometry \citep{Har-Tho:1968:ASTRJ2:SlowRotRelStarII} characterized by three parameters: mass $M$, internal angular momentum $J$ and quadrupole moment $Q$. It is convenient to use dimensionless parameters $a = J/M^2$ (spin) and $q = QM/J^2$ (dimensionless quadrupole moment). In the~special case when $q \sim 1$, the~Hartle--Thorne geometry reduces to the~well known and well studied Kerr geometry that is convenient for calculations in strong gravitational field regime because of high simplicity of relevant formulae. Near-maximum-mass neutron (quark) star Hartle--Thorne models constructed for any given equation of state imply $q \sim 1$ and the~Kerr geometry is applicable quite correctly in such situations instead of the~Hartle--Thorne geometry \citep[][Urbanec
et al. in preparation]{Tor-etal:2010:ASTRJ2:MassConstraints}. High neutron star masses can be expected in the~LMXB as a result of accretion and in such a case the~use of the~simple Kerr geometry is justified.

Quality of the~fitting procedure appears to be poor for the~atoll source 4U~1636$-$53 showing resonant frequency ratios 3\,:\,2 and 5\,:\,4 \citep{Tor-etal:2012:ApJ:}. Similar very bad fit of observational data to the~frequency-relation models was  found by \citet{Lin-etal:2011:ApJ:} for the~atoll sources 4U~1636$-$53 and Sco~X-1: the~mass and spin determined by the~fitting procedure are given with very large error for any of the~applied models used in the~paper, as also for some other models of the~frequency relations \citep{Mil-Lam-Psa:1998:ApJ:,Osh-Tit:1999:,Zha:2004:,Zha-etal:2006:MNRAS:,Muk:2009:ApJ:,Shi-Li:2009:,Chak-etal:2009:,Shi:2011:,Muk-Bha:2012:ApJ}. The~strong disagreement of the~data distribution and their fitting by the~frequency-relation models based on the~assumption of the~geodesic character of the~oscillatory frequencies related to proper combinations of the~geodetical orbital and epicyclic motion caused attempts to find a~correction of a~non-geodesic origin reflecting some important physical ingredients -- as e.g., influence of the~magnetic field of the~neutron star \citep{Bak-etal:2012:CQG:,Bak-etal:2010:CLAQG:MagIndNonGeo,Kov-Stu-Kar:2008:CLAQG:OffEqOrb,Stu-Kol:2012:JCAP:}, or of thickness of non-slender oscillating tori \citep{Rez-Yos-Zan:2003:MONNR:,Str-Sra:2009:CLAQG:EpiOscNonSleKerrBH} that could make the~fitting procedure much better, as shown in \citet{Tor-etal:2012:ApJ:}.

However, it is useful to consider another possibility to improve the~fitting procedures that is not based on the~non-geodesic corrections and modifies the~present orbital resonance models based on the~assumption of the~geodesic orbital and epicyclic motion, or disk oscillations with frequencies related to those of the~geodetical motion.

\section{Resonant switch model of HF~QPOs\\in neutron star systems}

We propose a~new model of twin peak HF~QPOs assuming that the~twin oscillatory modes creating sequences of the~lower and upper HF~QPOs can switch at a~resonant point. According to such a~Resonant Switch (RS) model non-linear resonant phenomena will cause excitation of a~new oscillatory mode (or two new oscillatory modes) and vanishing of one of the~previously acting modes (or both the~previous modes), i.e., switching from one pair of the~oscillatory modes to other pair of them that will be acting up to the~following resonant point. We assume two resonant points at the~disk radii $r_{\mathrm{out}}$ and $r_{\mathrm{in}}$, with observed frequencies $\nu_{\mathrm{U}}^{\mathrm{out}}$, $\nu_{\mathrm{L}}^{\mathrm{out}}$ and $\nu_{\mathrm{U}}^{\mathrm{in}}$, $\nu_{\mathrm{L}}^{\mathrm{in}}$, being in commensurable ratios
$p^{\mathrm{out}} = n^{\mathrm{out}}: m^{\mathrm{out}}$ and $p^{\mathrm{in}} = n^{\mathrm{in}}: m^{\mathrm{in}}$. These resonant frequencies are determined by the~energy switch effect \citep{Tor:2009:ASTRA:ReversQPOs}; observations put the~restrictions $\nu_{\mathrm{U}}^{\mathrm{in}} > \nu_{\mathrm{U}}^{\mathrm{out}}$ and $p^{\mathrm{in}} < p^{\mathrm{out}}$. In the~region covering the~resonant point at $r_{\mathrm{out}}$ the~twin oscillatory modes with the~upper (lower) frequency are determined by the~function $\nu_{\mathrm{U}}^{\mathrm{out}}(x;M,a)$ ($\nu_{\mathrm{L}}^{\mathrm{out}}(x;M,a)$). Near the~inner resonant point at $r_{\mathrm{in}}$ different oscillatory modes given by the~frequency functions $\nu_{\mathrm{U}}^{\mathrm{in}}(x;M,a)$ and $\nu_{\mathrm{L}}^{\mathrm{in}}(x;M,a)$ occur. We assume all the~frequency functions to be determined by combinations of the~orbital and epicyclic frequencies of the~geodesic motion in the~Kerr backgrounds. Such a~simplification is correct with high precision for neutron (quark) stars with large masses, close to maximum allowed for a~given equation of state. As demonstrated in \citet{Tor-etal:2010:ASTRJ2:MassConstraints} and Urbanec
et al. (in preparation), the~quadrupole moment of Hartle--Thorne geometry corresponding to near-extreme masses is very close to the~Kerr limit and the~orbital and epicyclic frequencies are very close to those given by the~exact Kerr geometry.

The~frequency functions have to meet the~observationally given resonant frequencies. In the~framework of the~simple RS~model, when two resonant points and two pairs of the~frequency functions are assumed, this requirement enables determination of the~parameters of the~Kerr background describing with high precision the~exterior of the~neutron (quark) star. The~``shooting'' of the~frequency functions to the~resonant points, giving the~neutron star parameters, can be realized efficiently in two steps. Independence of the~frequency ratio on the~mass parameter $M$ implies the~conditions
\begin{eqnarray}
      \nu_{\mathrm{U}}^{\mathrm{out}}(x;M,a) &:& \nu_{\mathrm{L}}^{\mathrm{out}}(x;M,a) = p^{\mathrm{out}}\,,\\
      \nu_{\mathrm{U}}^{\mathrm{in}}(x;M,a) &:& \nu_{\mathrm{L}}^{\mathrm{in}}(x;M,a) = p^{\mathrm{in}}
\end{eqnarray}
giving relations for the~spin $a$ in terms of the~dimensionless radius $x$ and the~resonant frequency ratio $p$. They can be expressed in the~form $a^{\mathrm{out}}(x,p^{\mathrm{out}})$ and $a^{\mathrm{in}}(x,p^{\mathrm{in}})$, or in an~inverse form $x^{\mathrm{out}}(a,p^{\mathrm{out}})$ and $x^{\mathrm{in}}(a,p^{\mathrm{in}})$. At the~resonant radii the~conditions
\begin{equation}
        \nu^{\mathrm{out}}_{\mathrm{U}} = \nu^{\mathrm{out}}_{\mathrm{U}}(x;M,a)\,, \quad    \nu^{\mathrm{in}}_{\mathrm{U}} = \nu^{\mathrm{in}}_{\mathrm{U}}(x;M,a)
\end{equation}
are satisfied along the~functions $M^{\mathrm{out}}_{p_{\mathrm{out}}}(a)$ and $M^{\mathrm{in}}_{p_{\mathrm{in}}}(a)$ which are obtained by using the~functions $a^{\mathrm{out}}(x,p^{\mathrm{out}})$ and $a^{\mathrm{in}}(x,p^{\mathrm{in}})$. The~parameters of the~neutron (quark) star are then given by the~condition
\begin{equation}
        M^{\mathrm{out}}_{p_{\mathrm{out}}}(a) = M^{\mathrm{in}}_{p_{\mathrm{in}}}(a)\,.    \label{RS}
\end{equation}
The~condition (\ref{RS}) determines $M$ and $a$ precisely, if the~resonant frequencies are determined precisely. If an~error occurs in determination of the~resonant frequencies, as naturally expected, our method gives related intervals of acceptable values of mass and spin parameter of the~neutron (quark) star.

Predictions of the~RS~model have to be tested. First, the~predicted ranges of the~mass and spin parameters have to be confronted with limits predicted by theoretical models of the~neutron (quark) star structure. Second, these ranges have to be tested by the~observational limits on mass and spin given by different phenomena observed in X-rays coming from the~source, e.g., by the~profiled spectral lines. Of course, crucial will be the test of fitting the twin peak HF QPO data around the resonant points with the results of the RS model.

Here we restrict ourselves to a~rough confrontation of the~neutron star theoretical predictions on the~mass and spin. The~other tests are under investigation and will be presented in a~forthcoming paper.

\section{Orbital and epicyclic frequencies of geodetical\\motion in the~Kerr geometry}

The~formulae for the~vertical epicyclic frequency $\nu_{\theta}$ and the~radial epicyclic frequency $\nu_{r}$ take in the~Kerr spacetime (describing black holes or exterior of near-maximum mass neutron stars) the~form  \citep[e.g.,][]{Ali-Gal:1981:GENRG2:,Kat-Fuk-Min:1998:BHAccDis:,Ste-Vie:1998:ASTRJ2L:,Ter-Stu:2005:ASTRA:}
\begin{equation}\label{frequencies}
\nu_{\theta}^2 = \alpha_\theta\,\nu_\mathrm{K}^2,
\qquad
\nu_{r}^2 = \alpha_{r}\,\nu_\mathrm{K}^2,
\end{equation}
where the~Keplerian orbital frequency $\nu_\mathrm{K}$ and the~related epicyclic frequencies are given by the~formulae
\begin{eqnarray}
\nu_{\mathrm{K}}&=&\frac{1}{2\pi}\left(\frac{\mathrm{G}M}{r_\mathrm{G}^{~3}}\right)^{1/2}\frac{1}{x^{3/2} + a} =
\frac{1}{2\pi}\left(\frac{\mathrm{c}^3}{\mathrm{G}M}\right)
\frac{1}{x^{3/2}+a}\,,\\
\alpha_\theta&=& 1-\frac{4\,a}{x^{3/2}}+\frac{3\,a^2}{x^{2}}\,,\\
\alpha_{r}&=&1-\frac{6}{x}+\frac{8\,a}{x^{3/2}}-\frac{3\,a^2}{x^{2}}\,.\label{alfa-r}
\end{eqnarray}
Here $x = r/(\mathrm{G}M/\mathrm{c}^2)$ is the~dimensionless radius, expressed in terms of the~gravitational radius.

The~Keplerian frequency $\nu_{\mathrm{K}}(x,a)$ is a~monotonically decreasing function of the~radial coordinate for any value of the~Kerr geometry spin. The~radial epicyclic frequency has a~global maximum for any Kerr black hole spacetime ($0<a<1$) and also the~vertical epicyclic frequency is not monotonic if the~spin is sufficiently high \citep[see, e.g.,][]{Kat-Fuk-Min:1998:BHAccDis:,Per-etal:1997:ASTRJ2:}. For the~Kerr black-hole spacetimes, the~locations $\mathcal{R}_{r}\,(a),~\mathcal{R}_\theta\,(a)$ of maxima of the~epicyclic frequencies $\nu_{r},~\nu_\theta$ are implicitly given by the~conditions \citep{Ter-Stu:2005:ASTRA:}
\begin{eqnarray}
\label{implicitcondition}
\beta_\mathrm{j}(x,a)&=&\frac{1}{2}\frac{\sqrt{x}}{x^{3/2}+a}\,\alpha_\mathrm{j}(x,a)\,,\quad \mathrm{where}\ ~\mathrm{j}\in\{{r},\theta\}\,,\\
\beta_{{r}}(x,a)&\equiv&\frac{1}{x^{2}}-\frac{2\,a}{x^{5/2}}+ \frac {a^2}
{x^3}\,,\\
\beta_{\theta}(x,a)&\equiv&\frac{a}{x^{5/2}}-\frac{a^2}{x^3}\,.
\end{eqnarray}
For any black hole spin, the~extrema of the~radial epicyclic frequency $\mathcal{R}_{r}\,(a)$ are located above the~marginally stable orbit. The~latitudinal epicyclic frequency has extrema $\mathcal{R}_\theta\,(a)$ located above the~photon (marginally bound or marginally stable) circular orbit in the~black hole spacetimes with spin $a>0.748$ (0.852, 0.952) \citep{Ter-Stu:2005:ASTRA:}. In the~Keplerian disks with the~inner boundary at $x_{\mathrm{in}} \sim x_{\mathrm{ms}}$, the~limiting value $a=0.952$ is relevant. Note that in the~Kerr naked singularity spacetimes (with $a>1$) the~behaviour of the~orbital and epicyclic frequencies is much more complex \citep{Ter-Stu:2005:ASTRA:,Stu-Sch:2012:CQG:ObsPhenKerrSSp:} and also the~related optical phenomena (as profiled lines)  demonstrate strong differences in comparison to those created in the~black hole or neutron star external spacetimes having $a<1$ \citep{Stu-Sch:2010:CLAQG:AppKepDiOrKerrSSp,Stu-Sch:2012:CQG:ObsPhenKerrSSp:}. However, these differences are important only at radii $r<M$, and become irrelevant at radii related to the~exterior of the~superspinning quark stars having (slightly) $a>1$ \citep{Lo-Lin:2011:ApJ:}.

\section{Frequency relations of the~oscillatory mode pairs}

We concentrate attention on the~combinations of the~relativistic precession (RP) and total precession (TP) frequency relations, defined by the~frequency mix for the~lower frequency, their modifications for the~upper frequency,  including the~higher, ``beat'', harmonics, and the~tidal disruption (TD) and warped  disk (WD) models. For each of the~frequency relations under consideration we present the~function determining the~resonant radii $x_{n:m}(a)$ given by the~upper to lower frequency ratio $\nu_{\mathrm{U}} : \nu_{\mathrm{L}} = n:m$ characterized by the~parameter
\begin{equation}
             p = \left(\frac{m}{n}\right)^2\, .
\end{equation}
Generally, the~condition $x_{n:m} > x_{\mathrm{ms}}$ has to be satisfied; the~marginally stable orbit $x_{\mathrm{ms}}$, assumed to correspond to the~inner edge of the~Keplerian disks, is implicitly determined by the~function
\begin{equation}
           a = a_{\mathrm{ms}}(x)\equiv\frac{\sqrt{x}}{3}\left(4-\sqrt{3x-2}\right).
\end{equation}

\subsection{Relativistic precession model}

The~RP model is determined by the~frequency relation
\begin{equation}
            \frac{\nu_{\mathrm{K}}}{\nu_{\mathrm{K}}-\nu_{{r}}}=\frac{n}{m}\,.
\end{equation}
The~resonance condition reads
\begin{equation}
           a=a^{\mathrm{K}/(\mathrm{K}-{r})}(x,p)\equiv \frac{\sqrt{x}}{3}\left[4-\sqrt{3x(1-p_{\mathrm{RP}})-2}\right]
\end{equation}
where
\begin{equation}
           p_{\mathrm{RP}}=\left(\frac{n-m}{n}\right)^2=\left(1-\sqrt{p}\right)^2.
\end{equation}

\textbf{Modification RP1.} The~upper frequency corresponds to the~vertical epicyclic frequency, the~modified frequency relation is determined by
\begin{equation}
            \frac{\nu_{\mathrm{\theta}}}{\nu_{\mathrm{K}}-\nu_{{r}}}=\frac{n}{m}\,.
\end{equation}
The~resonance function $a^{\theta/(\mathrm{K}-{r})}(x,p)$ is properly chosen solution of the~equation
\begin{equation}
            p^2\alpha_{\theta}^2-2p\alpha_{\theta}\left(1+\alpha_{{r}}\right)+\left(1-\alpha_{{r}}\right)^2=0\,.
\end{equation}
We shall not give the~function explicitly here because of its complexity.

\textbf{Modification RPB.} The~upper frequency corresponds to the~beat frequency, the~modified frequency relation is determined by
\begin{equation}
            \frac{\nu_{\mathrm{K}}+\nu_{{r}}}{\nu_{\mathrm{K}}-\nu_{{r}}}=\frac{n}{m}\,.
\end{equation}
The~resonance condition reads
\begin{equation}
            a=a^{(\mathrm{K}+{r})/(\mathrm{K}-{r})}(x,p)\equiv
            \frac{\sqrt{x}}{3}\left[4-\sqrt{3x(1-p_{\mathrm{RPB}})-2}\right]
\end{equation}
where
\begin{equation}
            p_{\mathrm{RPB}}=\left(\frac{n-m}{n+m}\right)^2=\left(\frac{\sqrt{p}-1}{\sqrt{p}+1}\right)^2.
\end{equation}

\subsection{Total precession model}

The~TP model is determined by the~frequency relation
\begin{equation}
            \frac{\nu_{\mathrm{\theta}}}{\nu_{\mathrm{\theta}}-\nu_{{r}}}=\frac{n}{m}\,.
\end{equation}
The~resonance condition reads
    \begin{eqnarray}
          a=a^{\theta/(\theta-r)}(x,p)&\equiv&\frac{\sqrt{x}}{3(p_{\mathrm{TP}}+1)}\Bigg\{2\left(p_{\mathrm{TP}}+2\right)-\nonumber\\
          && -\sqrt{(1-p_{\mathrm{TP}})\left[3x(p_{\mathrm{TP}}+1)-2(2p_{\mathrm{TP}}+1)\right]}\Bigg\}
    \end{eqnarray}
where
\begin{equation}
           p_{\mathrm{TP}}=\left(\frac{n-m}{n}\right)^2=\left(1-\sqrt{p}\right)^2.
\end{equation}

\textbf{Modification TP1.} The~upper frequency corresponds to the~Keplerian orbital frequency, the~modified frequency relation is determined by
\begin{equation}
            \frac{\nu_{\mathrm{K}}}{\nu_{\mathrm{\theta}}-\nu_{{r}}}=\frac{n}{m}\,.
\end{equation}
The~resonance function $a^{\mathrm{K}/(\theta-{r})}(x,p)$ is solution of the~equation
    \begin{equation}
       \left(\alpha_{\theta}-\alpha_{{r}}\right)^2-2p\left(\alpha_{\theta}+\alpha_{{r}}\right)+p^2=0\,.
    \end{equation}
       In the explicit form the resonance condition reads
     \begin{eqnarray}
      a&=&a^{\mathrm{K}/(\theta-{r})}(x,p) \equiv\nonumber\\
        &\equiv&\sqrt{x}+\frac{1}{2\cdot 3^{5/6}}\times\,\left[\sqrt{\frac{A^{2/3}+B}{A^{1/3}}}-\sqrt{A^{1/3} \left(\frac{4 \sqrt{3} p
        x^{5/2}}{\sqrt{A+A^{1/3} B}}-1\right)-\frac{B}{A^{1/3}}}\,\right]
    \end{eqnarray}
where
       \begin{eqnarray}
        A&=&6 p^2 x^5+\sqrt{36 p^4 x^{10}-B^3}\,,\\
        B&=&3^{1/3} p x^3 \left[4+(p-4) x\right]\,.
       \end{eqnarray}

\textbf{Modification TPB.} The~upper frequency corresponds to the~beat frequency, the~modified frequency relation is determined by
\begin{equation}
            \frac{\nu_{\theta}+\nu_{r}}{\nu_{\theta}-\nu_{r}}=\frac{n}{m}\,.
\end{equation}
The~resonance condition reads
      \begin{eqnarray}
         a=a^{(\theta + r)/(\theta-r)}(x,p)&\equiv&
          \frac{\sqrt{x}}{3(p_{\mathrm{TPB}}+1)}\Bigg\{2(p_{\mathrm{TPB}}+2)-\nonumber\\
          && -\sqrt{(1-p_{\mathrm{TPB}})\left[3x(p_{\mathrm{TPB}}+1)-2(2p_{\mathrm{TPB}}+1)\right]}\Bigg\}
      \end{eqnarray}
where
\begin{equation}
           p_{\mathrm{TPB}}=\left(\frac{n-m}{n+m}\right)^2=\left(\frac{\sqrt{p}-1}{\sqrt{p}+1}\right)^2 .
\end{equation}

\subsection{Tidal disruption model}

The~TD model \citep{Kos-etal:2009:tidal:} is based on the~idea of an~orbiting hot spot distorted by the~influence of the~tidal effects of the~central black hole or neutron star. It is is determined by the~frequency relation
\begin{equation}
                    \frac{\nu_{\mathrm{K}}+\nu_{{r}}}{\nu_{\mathrm{K}}}=\frac{n}{m}\,.
\end{equation}
The~resonance condition reads
\begin{equation}
            a=a^{(\mathrm{K}+{r})/\mathrm{K}}(x,p)\equiv
            \frac{\sqrt{x}}{3}\left[4-\sqrt{3x(1-p_{\mathrm{TD}})-2}\right]
\end{equation}
where
\begin{equation}
            p_{\mathrm{TD}}=\left(\frac{n-m}{m}\right)^2=\frac{\left(1-\sqrt{p}\right)^2}{p}\,.
\end{equation}

\subsection{Warped disk model}

The~warped disk oscillations (WD) model \citep{Kat:2004:PUBASJ:QPOsmodel,Kato:2008:b:PASJ:,Wag:1999:Discoseismology} assumes the~frequency of the~disk oscillations to be given by combinations of the~(multiples) of the~Keplerian and epicyclic frequencies. Usually, the~inertial-acoustic and g-mode oscillations and their resonances could be relevant. We consider as an~example the~frequency relation
\begin{equation}
    \frac{2\nu_{\mathrm{K}}-\nu_{r}}{2\left(\nu_{\mathrm{K}}-\nu_{r}\right)}=\frac{n}{m}\,.
\end{equation}
The~resonance condition reads
\begin{equation}
           a=a^{(2\mathrm{K}-r)/(2{\mathrm{K}}-2r)}(x,p)\equiv \frac{\sqrt{x}}{3}\left[4-\sqrt{3x\left(1-p_{\mathrm{WD}}\right)-2}\right]
\end{equation}
where
\begin{equation}
           p_{\mathrm{WD}}=\left[\frac{2(n-m)}{2n-m}\right]^2=\left[\frac{2\left(1-\sqrt{p}\right)}{2-\sqrt{p}}\right]^2.
\end{equation}

\section{Resonant switch model applied to the~source\\4U~1636$-$53}

The~observational X-ray data obtained for both the~atoll and Z-sources indicate feasibility of the~RS~model at least in some of the~observed sources. We shall test the~atoll source 4U~1636$-$53, where the~twin peak HF~QPOs span a~wide range of frequencies crossing the~frequency ratios 3\,:\,2, 4\,:\,3 and finishing at 5\,:\,4, just near (or at) the~inner edge of the~accretion disk \citep{Bou-Bar-Lin-Tor:2010:MNRAS:,Tor:2009:ASTRA:ReversQPOs,Stu-Kot-Tor:2011:ASTRA:ResRadKep}. In the~case of the~4U~1636$-$53 source, behaviour of the~observed oscillations  indicates presence of the~resonant effects at disk radii where the~frequencies are in the~ratios 3\,:\,2 and 5\,:\,4 because of the~energy switch effect occurring at these frequency ratios \citep{Tor:2009:ASTRA:ReversQPOs}; therefore, the~source is properly chosen for our test of the~RS~model.

\subsection{The~resonant frequencies}
Using the~results of \citet{Tor:2009:ASTRA:ReversQPOs}, the~resonant frequencies determined by the~energy switch effect are given in the~outer resonant point with frequency ratio $\nu_{\mathrm{U}}/\nu_{\mathrm{L}}=3/2$ by the~frequency intervals
\begin{eqnarray}
        \nu_{\mathrm{U}}^{\mathrm{out}} &=& \nu_{\mathrm{U0}}^{\mathrm{out}} \pm \Delta \nu^{\mathrm{out}} = (970 \pm 30) \,\mathrm{Hz}\,,\nonumber\\
        \nu_{\mathrm{L}}^{\mathrm{out}} &=& \nu_{\mathrm{L0}}^{\mathrm{out}} \pm \Delta \nu^{\mathrm{out}} = (647 \pm 20) \,\mathrm{Hz}\,,  \label{freu}
\end{eqnarray}
while at the~inner resonant point with frequency ratio $\nu_{\mathrm{U}}/\nu_{\mathrm{L}}=5/4$ the~frequency intervals are
\begin{eqnarray}
        \nu_{\mathrm{U}}^{\mathrm{in}} &=& \nu_{\mathrm{U0}}^{\mathrm{in}} \pm \Delta \nu^{\mathrm{in}} =  (1180 \pm 20) \,\mathrm{Hz}\,,\nonumber\\
        \nu_{\mathrm{L}}^{\mathrm{in}} &=& \nu_{\mathrm{L0}}^{\mathrm{in}} \pm \Delta \nu^{\mathrm{in}} =   (944 \pm 16) \,\mathrm{Hz}\,. \label{frei}
\end{eqnarray}
Note that the~resonant points determined by using the~energy switch effect are in accord with observational data points crossing the~lines of constant frequency ratios 3\,:\,2 and 5\,:\,4 as given in the~standard papers \citep{Bar-Oli-Mil:2005:MONNR:,Bel-etal:2007:MONNR:RossiXTE}.

\subsection{Theoretical limits on the~mass and spin of neutron\\and quark stars}

It is well known that the~neutron star mass surely cannot exceed the~critical  value of $M \sim 3.2\,\mathrm{M}_{\odot}$ \citep{Rho-Ruf:1974:PHYRL:}. On the~other hand, realistic equations of state put limit on the~maximal mass of neutron stars around $M_{\mathrm{maxN}} \sim 2.8\,\mathrm{M}_{\odot}$ \citep{Pos-etal:2010:LoveNum:}. In fact, the~extremal maximum $M_{\mathrm{maxN}} \sim 2.8\,\mathrm{M}_{\odot}$ is predicted by the~field theory \citep{Mul-Ser:1996:}. The~limit of $M_{\mathrm{maxN}} \sim 2.5\,\mathrm{M}_{\odot}$ is predicted by Dirac--Brueckner--Hartree--Fock approach in some special case \citep{Mut-etal:1987:} and the~variational approaches \citep{Akm-Pan:1997:,Akm-Pan:1998:} and other approaches \citep{Urb-Bet-Stu:2010:ACTA:ObsTestNSMeanField} allow for $M_{\mathrm{maxN}} \sim 2.25\,\mathrm{M}_{\odot}$.
On the~neutron star dimensionless spin the~limit of $a< a_{\mathrm{maxN}}=0.7$ has been recently reported, being  independent of the~equation of state \citep{Lo-Lin:2011:ApJ:}.

For the~quark stars the~maximal mass is expected somewhat smaller in comparison with the~neutron stars because of softer equations of state assumed in modelling the~quark stars, but masses around $M_{\mathrm{maxQ}} \sim 2\,\mathrm{M}_{\odot}$ are allowed \citep[see, e.g.,][]{Gle:2000:CompactStars:,Lo-Lin:2011:ApJ:}. However, a~substantial difference occurs in the~limit on maximal spin, since even slightly superspinning states of quark stars with $a_{\mathrm{maxQ}} \geq 1$,  exceeding the~black hole limit, has been recently reported by \citet{Lo-Lin:2011:ApJ:}, independently of the~details of the~equation of state for quark matter. Such a~sharp difference between the~limits on the~maximal spin of neutron and quark stars could probably be explained by the~fact that in quark (strange) stars the~strong nuclear force helps in binding the~stars (or could be the~binding force in low mass strange stars), while no such effect can be present in the~case of neutron stars.

In the~Hartle--Thorne models of rotating neutron stars the~spin of the~star has to be related to its rotational frequency. The~rotational frequency of the~neutron star at the~atoll source 4U~1636$-$53 has been observed at $f_{\mathrm{rot}} = 580\,\mathrm{Hz}$, or $f_{\mathrm{rot}} = 290\,\mathrm{Hz}$, if we observe doubled radiating structure \citep{Str-Mar:2002:ApJ:}. Of course, such a~rotational frequency is substantially lower in comparison with the~mass shedding frequency, and the~Hartle--Thorne model can be applied quite well, predicting spins significantly lower than the~maximally allowed spin.

The~theory of neutron star structure then implies for a~wide variety of realistic equations of state the~spin in the~range \citep[][Urbanec
et al. in preparation]{Har-Tho:1968:ASTRJ2:SlowRotRelStarII} 
\begin{equation}\label{limit-na-spin-z-rotace}
    0.1 < a < 0.4\,.
\end{equation}
Of course, the~upper part of the~allowed spin range corresponds to the~rotational frequency $f_{\mathrm{rot}} = 580\,\mathrm{Hz}$, while the~lower part corresponds to $f_{\mathrm{rot}} = 290\,\mathrm{Hz}$.
The~related restriction on the~neutron star (near-extreme) mass reads
\begin{equation}\label{limit-na-M-z-rotace}
    M<2.5\,\mathrm{M}_{\odot}\,.
\end{equation}

Therefore, we shall use these restrictions and make a~short comment on the~results of the~RS~model in the~case of three typical combinations considered in our study. We plan for a~future work to make a~detailed study considering Hartle--Thorne models constructed for a~large variety of equations of state allowing for sufficiently large mass of the~4U~1636$-$53 neutron (or quark) star.\footnote{Notice that quite recent study of the~HF~QPOs and line profiles observed in 4U~1636$-$53 leads the~authors to claim that combining these two effects they are able to predict mass of the~neutron star to be $\sim 2.4\,\mathrm{M}_{\odot}$ \citep{Sanna-etal-poster-IAU-Peking-2012}.}


\begin{figure}[t]
\begin{minipage}[h]{.49\hsize}
\begin{center}
\includegraphics[width=\hsize]{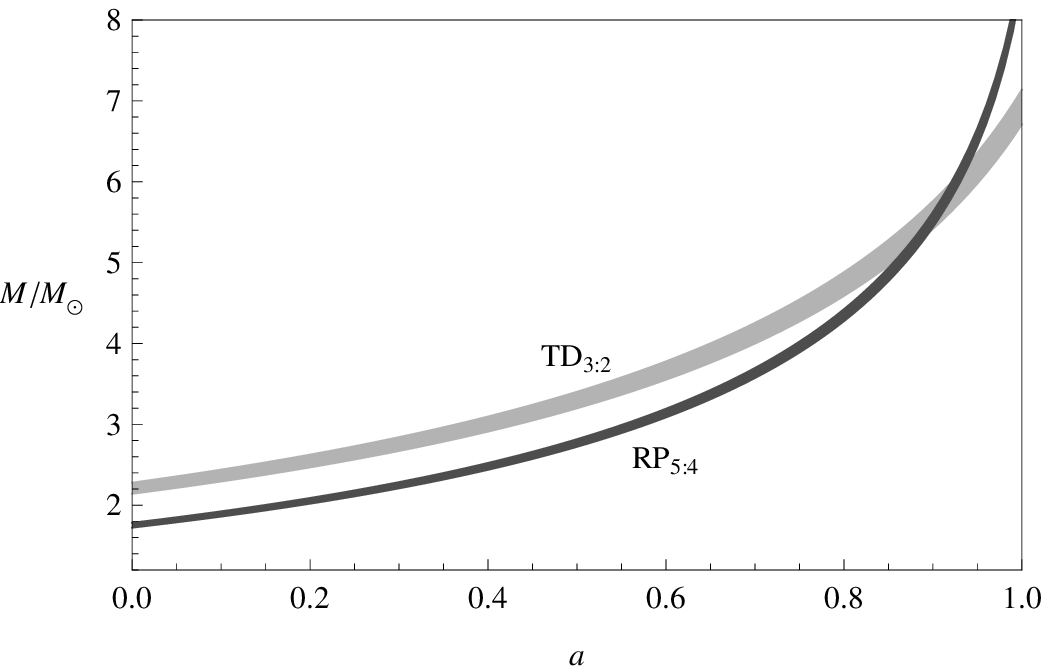}
\end{center}
\end{minipage}
\hfill
\begin{minipage}[h]{.49\hsize}
\begin{center}
\includegraphics[width=\hsize]{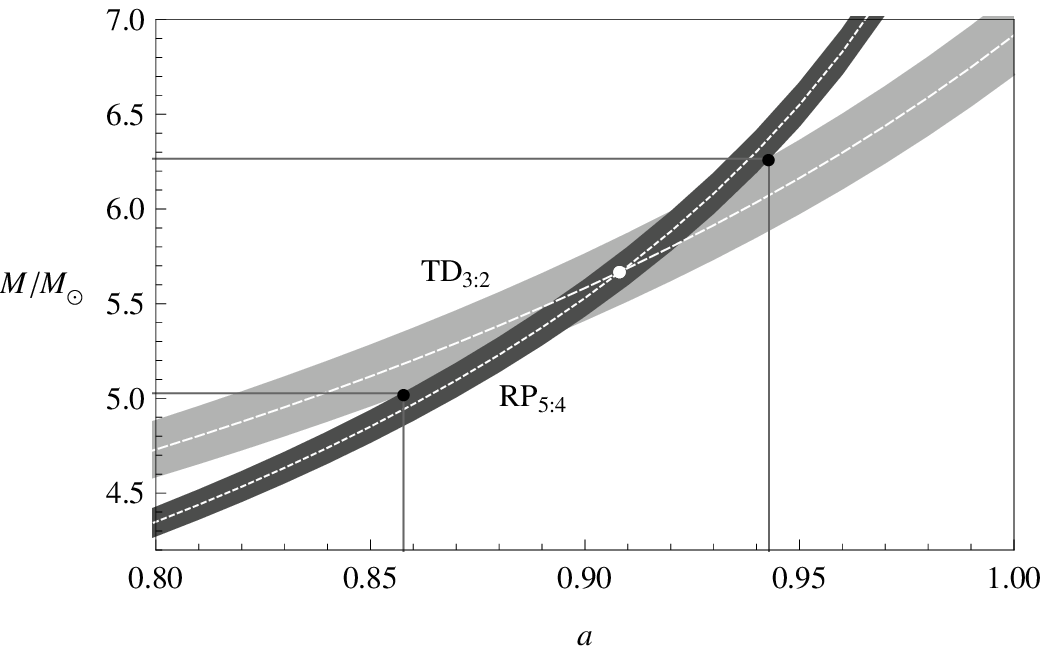}
\end{center}
\end{minipage}
\caption{\small \label{grafy-M-a-TD-RP}\textit{Left panel}: possible combinations of mass and spin of neutron star in the~atoll source 4U~1636$-$53 predicted by the~RP and TD model using the~scatter of the~resonant frequencies at each of the~two resonant points where $\mathrm{RP} \equiv \nu_\mathrm{K}:(\nu_\mathrm{K}-\nu_{r})=5:4$ and $\mathrm{TD} \equiv (\nu_\mathrm{K}+\nu_{r}):\nu_\mathrm{K} = 3:2$. The~\textit{right panel} gives the~detailed information about the~intervals of mass and spin relevant for this combination of the~two frequency relations assuming the~resonant switch model.}
\end{figure}

\begin{figure}[h]
\begin{minipage}[h]{.49\hsize}
\begin{center}
\includegraphics[width=\hsize]{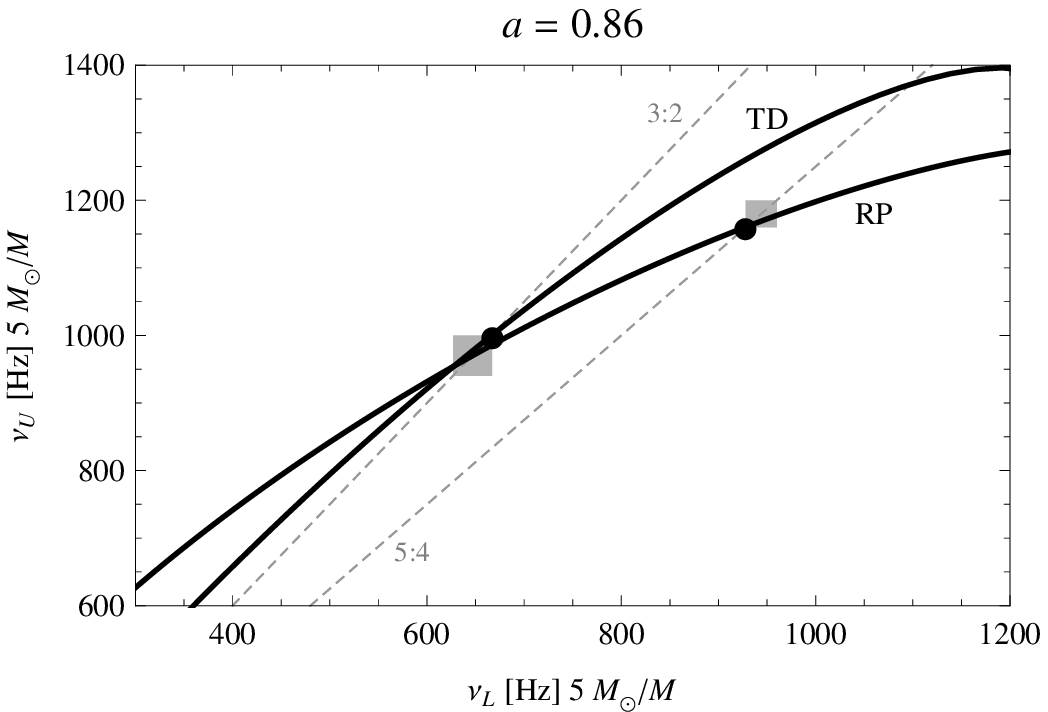}
(a)
\end{center}
\end{minipage}
\hfill
\begin{minipage}[h]{.49\hsize}
\begin{center}
\includegraphics[width=\hsize]{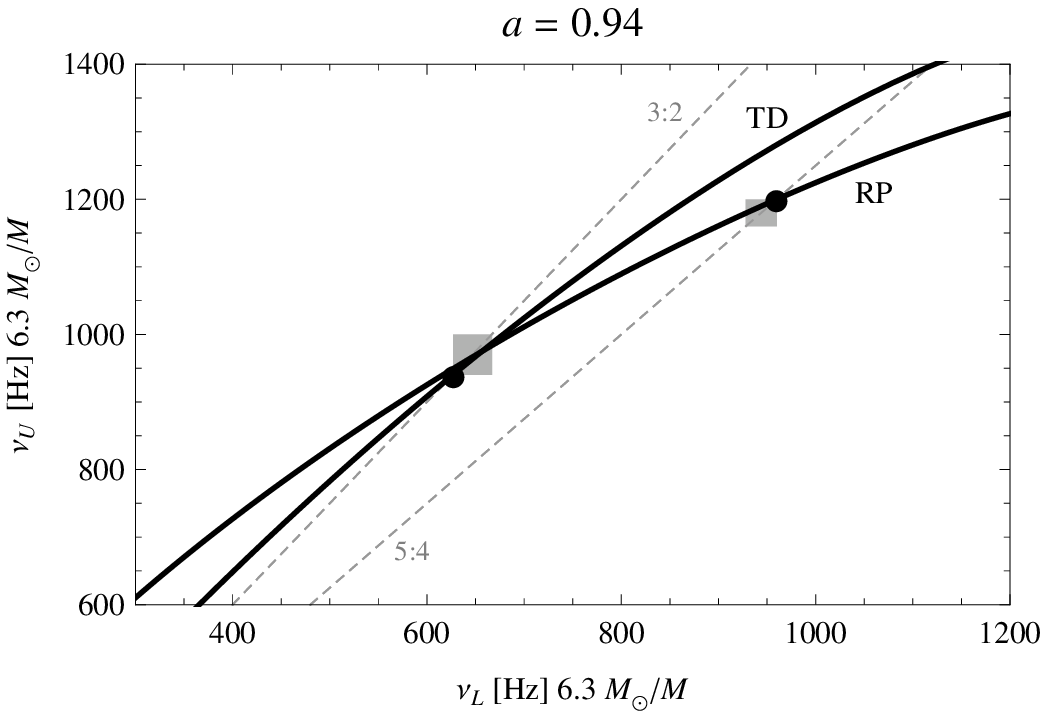}\\
(b)
\end{center}
\end{minipage}
\mbox{}
\begin{center}
\begin{minipage}[h]{.49\hsize}
\begin{center}
\includegraphics[width=\hsize]{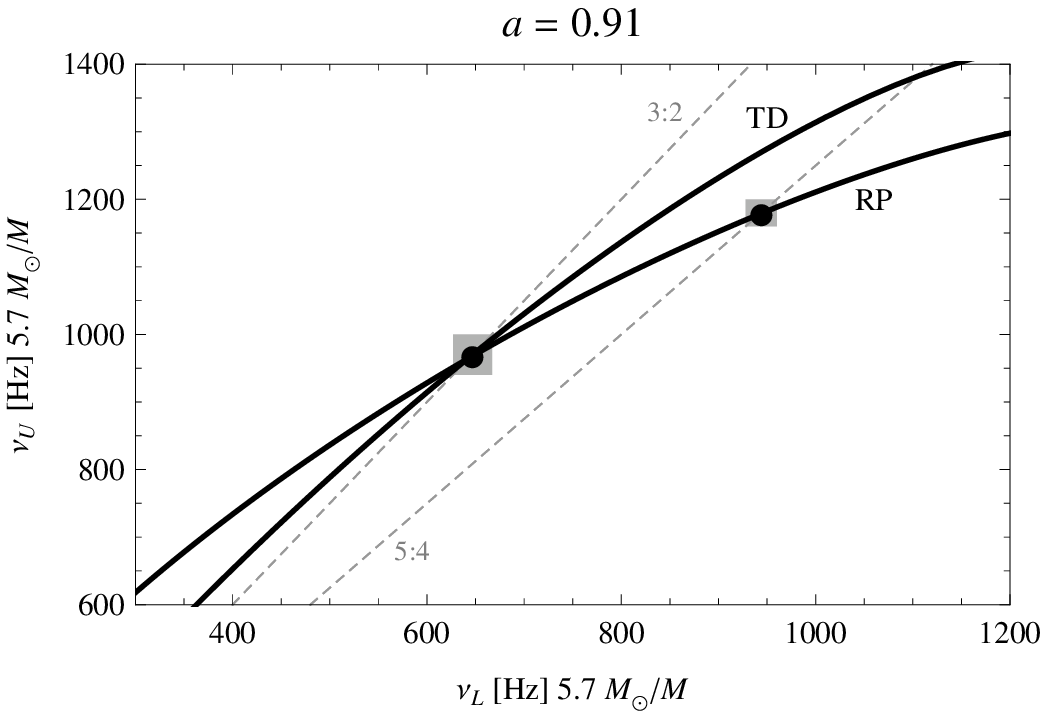}
(c)
\end{center}
\end{minipage}
\end{center}
\caption{\small RP and TD model fits for the~atoll source 4U~1636$-$53. The~gray rectangles indicate the~scatter of the~resonant frequencies at each of the~two resonant points. The~fits are shown for the~minimal \textit{(a)}, maximal \textit{(b)} and central \textit{(c)} allowed values of neutron star mass and spin implied by the~resonant switch model. The~case \textit{(a)} corresponds to the~left black point on the~\textit{right panel} of Fig.~\ref{grafy-M-a-TD-RP} with $\nu_{\mathrm{U}}^{\mathrm{out}}(3\!:\!2) = 1000\,\mathrm{Hz}$, $\nu_{\mathrm{L}}^{\mathrm{out}}(3\!:\!2) = 667\,\mathrm{Hz}$ and $\nu_{\mathrm{U}}^{\mathrm{in}}(5\!:\!4) = 1160\,\mathrm{Hz}$, $\nu_{\mathrm{L}}^{\mathrm{in}}(5\!:\!4) = 928\,\mathrm{Hz}$, \textit{(b)} corresponds to the~right black point on the~\textit{right panel} of Fig.~\ref{grafy-M-a-TD-RP} with $\nu_{\mathrm{U}}^{\mathrm{out}}(3\!:\!2)  = 940\,\mathrm{Hz}$, $\nu_{\mathrm{L}}^{\mathrm{out}}(3\!:\!2)  = 627\,\mathrm{Hz}$ and $\nu_{\mathrm{U}}^{\mathrm{in}}(5\!:\!4) = 1200\,\mathrm{Hz}$, $\nu_{\mathrm{L}}^{\mathrm{in}}(5\!:\!4) = 960\,\mathrm{Hz}$, and \textit{(c)} corresponds to the~central white point on the~\textit{right panel} of Fig.~\ref{grafy-M-a-TD-RP} with $\nu_{\mathrm{U}}^{\mathrm{out}}(3\!:\!2)  = 970\,\mathrm{Hz}$, $\nu_{\mathrm{L}}^{\mathrm{out}}(3\!:\!2)  = 647\,\mathrm{Hz}$ and $\nu_{\mathrm{U}}^{\mathrm{in}}(5\!:\!4) = 1180\,\mathrm{Hz}$, $\nu_{\mathrm{L}}^{\mathrm{in}}(5\!:\!4) = 944\,\mathrm{Hz}$.\label{fity-TD-RP}}
\end{figure}

\subsection{Ranges of 4U~1636$-$53 neutron star mass and spin implied\\by the~RS~model}

We have presented above a~variety of the~oscillatory mode pairs that deserve consideration and determined the~representative resonance functions $a^{\nu_\mathrm{U}/\nu_\mathrm{L}}(x,p)$ related to the~oscillatory pairs. Now we shall test all the~possible combinations of the~oscillatory pairs at the~established resonant points and give the~related intervals of allowed values of the~mass and spin of the~neutron or quark star. We shall discuss in detail two characteristic cases of the~combinations RP--TD, RP--TP and its modification RP1--TP1. In the~other cases, we shall give only the~resulting allowed intervals of the~neutron star parameters $M$ and $a$. Of course, each pair of the~frequency relations under consideration has to be properly ordered at the~resonant points.

We have made the~mass and spin estimates by ``shooting'' the~frequency relations to the~two resonant points for all the~frequency relations presented above. The~procedure of determining the~intervals of mass and spin relevant for the~combinations of the~frequency relations is demonstrated in Figs.~\ref{grafy-M-a-TD-RP}\,--\,\ref{fity-TP-RP}. The~intervals of the~mass and spin implied by the~procedure of the~RS~model are summarized in Table~\ref{Tab-M-a-pro-RPaTPmodely}. Only combinations giving a~RS~solution are presented in Table~\ref{Tab-M-a-pro-RPaTPmodely} where for the~frequency relations under consideration the~relevant resonant points are explicitly given.

\begin{figure}[t]
\begin{minipage}[h]{.49\hsize}
\begin{center}
\includegraphics[width=\hsize]{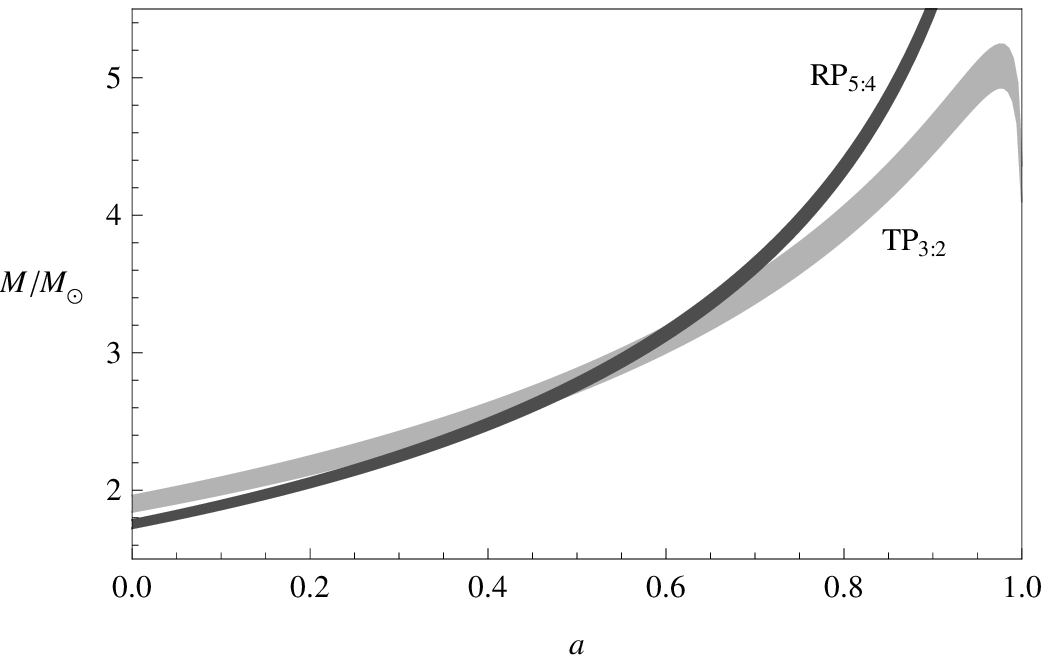}
\end{center}
\end{minipage}
\hfill
\begin{minipage}[h]{.49\hsize}
\begin{center}
\includegraphics[width=\hsize]{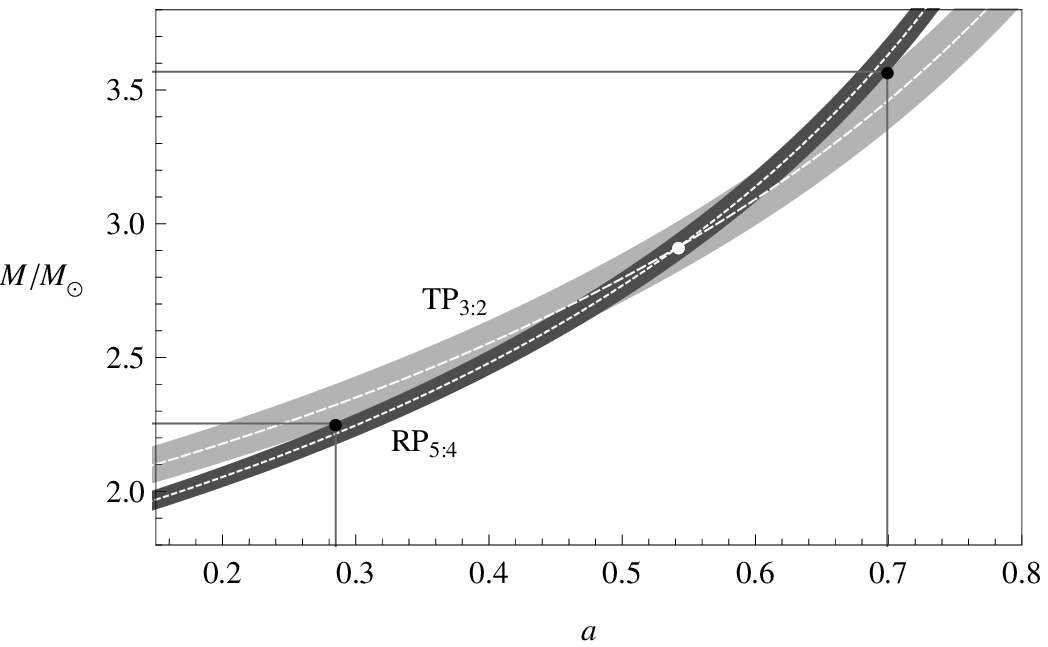}
\end{center}
\end{minipage}
\medskip
\caption{\small \label{grafy-M-a-TP-RP}\textit{Left panel}: possible combinations of mass and spin of neutron star in the~atoll source 4U~1636$-$53 predicted by the~RP and TP model using the~scatter of the~resonant frequencies at each of the~two resonant points where $\mathrm{RP} \equiv \nu_\mathrm{K}:(\nu_\mathrm{K}-\nu_{r})=5:4$ and $\mathrm{TP} \equiv \nu_\theta :(\nu_\theta-\nu_{r}) = 3:2$. The~\textit{right panel} gives the~detailed information about the~intervals of mass and spin relevant for this combination of the~two frequency relations assuming the~resonant switch model.}
\end{figure}

\begin{figure}[h]
\begin{minipage}[h]{.49\hsize}
\begin{center}
\includegraphics[width=\hsize]{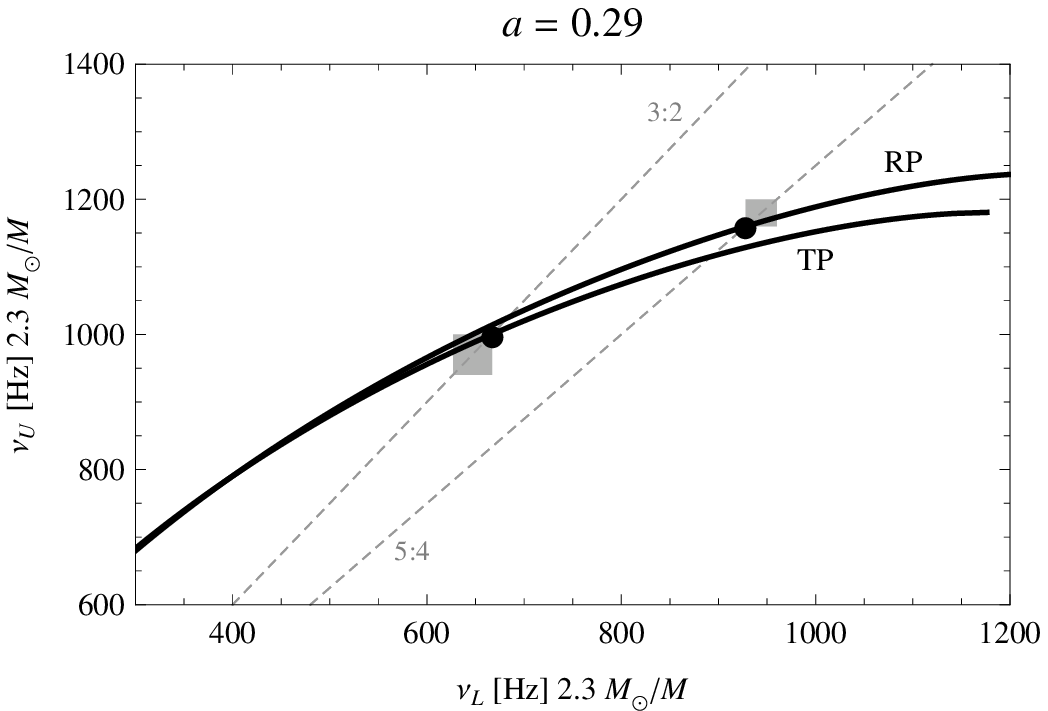}
(a)
\end{center}
\end{minipage}
\hfill
\begin{minipage}[h]{.49\hsize}
\begin{center}
\includegraphics[width=\hsize]{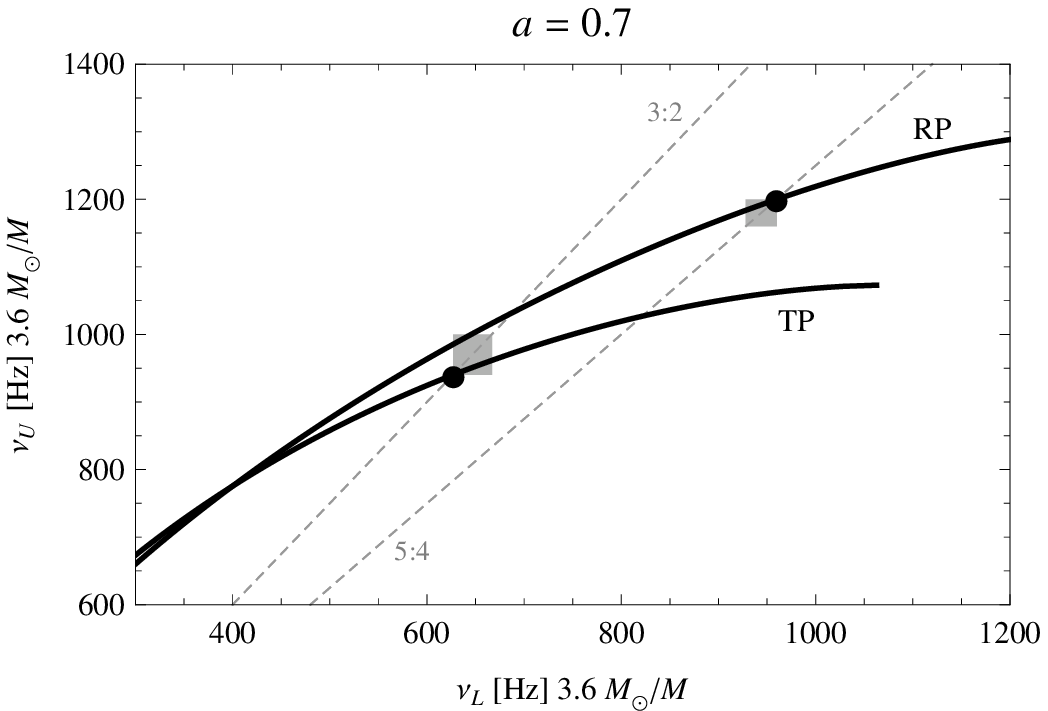}\\
(b)
\end{center}
\end{minipage}
\mbox{}
\begin{center}
\begin{minipage}[h]{.49\hsize}
\begin{center}
\includegraphics[width=\hsize]{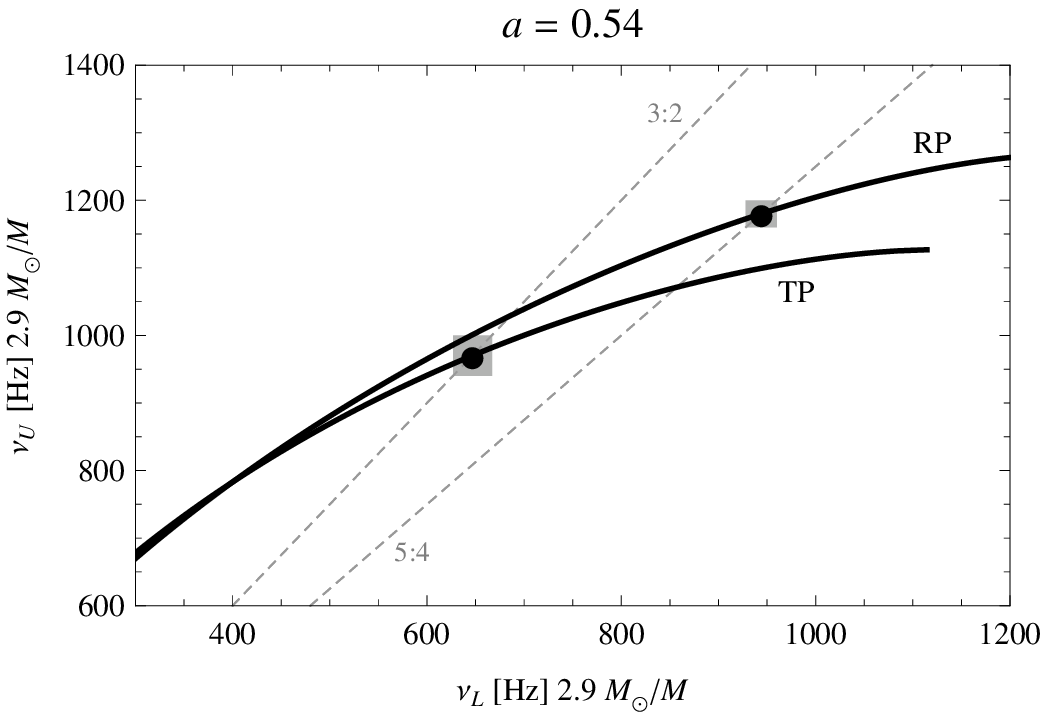}
(c)
\end{center}
\end{minipage}
\end{center}
\caption{\small RP and TP model fits for the~atoll source 4U~1636$-$53. The~gray rectangles indicate the~scatter of the~resonant frequencies at each of the~two resonant points. The~fits are shown for the~minimal \textit{(a)}, maximal \textit{(b)} and central \textit{(c)} allowed values of neutron star mass and spin implied by the~resonant switch model. The~case \textit{(a)} corresponds to the~left black point on the~\textit{right panel} of Fig.~\ref{grafy-M-a-TP-RP} with $\nu_{\mathrm{U}}^{\mathrm{out}}(3\!:\!2) = 1000\,\mathrm{Hz}$, $\nu_{\mathrm{L}}^{\mathrm{out}}(3\!:\!2) = 667\,\mathrm{Hz}$ and $\nu_{\mathrm{U}}^{\mathrm{in}}(5\!:\!4) = 1160\,\mathrm{Hz}$, $\nu_{\mathrm{L}}^{\mathrm{in}}(5\!:\!4) = 928\,\mathrm{Hz}$, \textit{(b)} corresponds to the~right black point on the~\textit{right panel} of Fig.~\ref{grafy-M-a-TP-RP} with $\nu_{\mathrm{U}}^{\mathrm{out}}(3\!:\!2)  = 940\,\mathrm{Hz}$, $\nu_{\mathrm{L}}^{\mathrm{out}}(3\!:\!2)  = 627\,\mathrm{Hz}$ and $\nu_{\mathrm{U}}^{\mathrm{in}}(5\!:\!4) = 1200\,\mathrm{Hz}$, $\nu_{\mathrm{L}}^{\mathrm{in}}(5\!:\!4) = 960\,\mathrm{Hz}$, and \textit{(c)} corresponds to the~central white point on the~\textit{right panel} of Fig.~\ref{grafy-M-a-TP-RP} with $\nu_{\mathrm{U}}^{\mathrm{out}}(3\!:\!2)  = 970\,\mathrm{Hz}$, $\nu_{\mathrm{L}}^{\mathrm{out}}(3\!:\!2)  = 647\,\mathrm{Hz}$ and $\nu_{\mathrm{U}}^{\mathrm{in}}(5\!:\!4) = 1180\,\mathrm{Hz}$, $\nu_{\mathrm{L}}^{\mathrm{in}}(5\!:\!4) = 944\,\mathrm{Hz}$.}\label{fity-TP-RP}
\end{figure}

The~mass and spin estimates of the~RS~model related to the~data of the~4U~1636-53 source have to be confronted with restrictions on the~neutron star mass and spin implied by the~theoretical models of neutron stars (or quark stars) and the~observed rotational frequency of the~neutron star at 4U~1636$-$53 presented above.

\subsubsection{Combination of the~RP and TD models}

For the~combination of RP and TD models, the~RP model has to be related to the~inner resonant point, while the~TD model has to be related to the~outer resonant point. We have demonstrated that for the~frequency scatter at the~resonant points, given by the~Eqs.~(\ref{freu}) and (\ref{frei}), the~range of allowed values of the~mass and spin of the~neutron star is given by
\begin{equation}
             0.86 < a < 0.94\, ,\quad
             5.03 < \frac{M}{\mathrm{M}_{\odot}} < 6.27 \,.
\end{equation}
The~central point of the~frequency ranges implies the~central estimates of the~neutron star parameters to be $a=0.91$ and $M = 5.67\,\mathrm{M}_{\odot}$. For completeness, we present also the~frequency dependence of the~RP and TD oscillatory modes for the~limiting values of the~spin and mass and for the~central values in Fig.~\ref{fity-TD-RP}. Qualitatively, these frequency dependencies could be in accord with the~observational data.

Clearly, in the~case of the~combination of the~RP and TD oscillatory modes the~RS model implies parameters of the~neutron star that are totally out of the~ranges accepted by the~recent theory of the~structure of the~neutron or quark stars both for their spin and mass. Therefore, the~RP--TD combination can be excluded as a~realistic explanation of the~observed data in the~atoll source 4U~1636$-$53 because of high values of the~predicted neutron star parameters. Nevertheless, it is interesting to notice that the~RP model fits in the~allowed range of mass and spin both resonant points quite well, especially for the~central point with $a=0.91$  and $M=5.7\,\mathrm{M}_{\odot}$, while the~TD frequency relation is restricted just to the~region corresponding to the~outer resonant point.

\renewcommand{\arraystretch}{1.15}
\begin{table}[!h]
\caption{{\small Intervals of mass and spin of the~neutron star in the~atoll source 4U~1636$-$53 implied by the~procedure of the~RS~model. The~shaded rows represent those
combinations of models that give acceptable values of spin and mass. Note that the~model WD(3:2) gives identical solution as TD(3:2).}\label{Tab-M-a-pro-RPaTPmodely}}
\begin{center}
\begin{tabular}{lcc}
  \hline
     \textbf{Combination of models} & \textbf{spin $a$} & \textbf{mass $[M/\mathrm{M}_{\odot}]$} \\
    \hline
RP(3:2) -- RP(5:4)  & $0.68 - 0.97$ & $3.60 - 6.88$ \\
\LCC \seda & \seda & \seda \\
 RP(5:4) -- RP1(3:2)  & $0.14 - 0.42$ & $1.98 - 2.49$ \\\ECC
 \LCC \seda & \seda & \seda \\
 RP(5:4) -- TP(3:2)   & $0.29 - 0.70$ & $2.25 - 3.57$ \\\ECC
 \LCC \seda & \seda & \seda \\
 RP(3:2) -- TP1(5:4)  & $0.27 - 0.74$ & $2.28 - 4.20$ \\\ECC
 \LCC \seda & \seda & \seda \\
 RP(3:2) -- TPB(5:4)  & $0.18 - 0.65$ & $2.24 - 3.45$ \\\ECC
 RP(5:4) -- TPB(3:2)  & $0.90 - 0.94$ & $5.63 - 6.27$ \\
\LCC \seda & \seda & \seda \\
 RP1(3:2) -- TP(5:4)  & $0.25 - 0.67$ & $2.11 - 2.94$ \\\ECC
 RP1(3:2) -- TP(5:4) & $0.988 - 0.994$ & $3.64 - 3.85$ \\
\LCC \seda & \seda & \seda \\
 RP1(3:2) -- TP1(5:4)  & $0.10 - 0.34$ & $1.94 - 2.36$ \\\ECC
 RP1(3:2) -- TPB(5:4)  & $0.992 - 0.996$ & $3.65 - 3.87$ \\
 RP1(5:4) -- TPB(3:2)  & $0.9993 - 0.9996$ & $3.47 - 3.59$ \\
\LCC \seda & \seda & \seda \\
 RPB(5:4) -- TP1(3:2)  & $0.18 - 0.67$ & $2.30 - 4.11$ \\\ECC
 RPB(5:4) -- TPB(3:2)  & $0.72 - 0.84$ & $4.52 - 5.56$ \\
\LCC \seda & \seda & \seda \\
 TP(3:2) -- TP1(5:4)  & $0.17 - 0.52$ & $2.07 - 2.93$ \\\ECC
 TP(3:2) -- TPB(5:4)  & $0.48 - 0.94$ & $2.84 - 4.71$ \\
\LCC \seda & \seda & \seda \\
 TP1(3:2) -- TPB(5:4)  & $0.08 - 0.38$ & $2.10 - 2.70$ \\\ECC
\LCC \seda & \seda & \seda \\
 TD(3:2) -- TD(5:4)   & $0.00 - 0.69$ & $2.14 - 4.22$ \\\ECC
 TD(3:2) -- RP(5:4)   & $0.86 - 0.94$ & $5.03 - 6.27$ \\
 TD(3:2) -- RPB(5:4)   & $0.34 - 0.73$ & $2.75 - 4.43$ \\
 TD(3:2) -- TP1(5:4)   & $0.70 - 0.74$ & $3.99 - 4.33$ \\
 TD(5:4) -- TP1(3:2)   & $0.38 - 0.71$ & $2.89 - 4.48$ \\
 TD(5:4) -- TPB(3:2)   & $0.70 - 0.85$ & $4.38 - 5.58$ \\
 WD(3:2) -- RP(5:4) & $0.86-0.94$ & $5.03 - 6.27$ \\
 WD(3:2) -- RPB(5:4)  & $0.34 - 0.73$ & $2.75 - 4.43$ \\
 WD(5:4) -- RPB(3:2)  & $0.00 - 0.86$ & $2.56 - 7.02$ \\
 WD(3:2) -- TP1(5:4)  & $0.70 - 0.74$ & $3.99 - 4.33$ \\
 WD(5:4) -- TP1(3:2)  & $0.85 - 0.89$ & $6.64 - 7.59$ \\
 WD(5:4) -- TPB(3:2)  & $0.00 - 0.29$ & $2.56 - 3.21$ \\
\LCC \seda & \seda & \seda \\
 WD(3:2) -- TD(5:4)   & $0.00 - 0.69$ & $2.14 - 4.22$ \\\ECC
 \hline
  \end{tabular}
\end{center}
\end{table}

\subsubsection{Combination of the~RP and TP models}

For the~combination of RP and TP models, the~RP model has to be related to the~inner resonant point, while the~TP model has to be related to the~outer resonant point. For the~frequency scatter at the~resonant points, given by the~Eqs.~(\ref{freu}) and (\ref{frei}), we deduce the~range of allowed values of the~mass and spin of the~neutron star given by
\begin{equation}
             0.29 < a < 0.70 \,,\quad
             2.25 < \frac{M}{\mathrm{M}_{\odot}} < 3.57\, .
\end{equation}
The~central point of the~frequency ranges implies the~central estimates of the~neutron star parameters to be $a=0.54$ and $M = 2.91\,\mathrm{M}_{\odot}$. In Fig.~\ref{fity-TP-RP} we present again the~frequency dependence of the~RP and TP oscillatory modes for the~limiting values of the~neutron star spin and mass and for the~central values of these parameters. Again, these frequency dependencies are in accord with the~observational data qualitatively. As in the~case of TD--RP combination, the~RP frequency relation meets both resonant points frequency intervals, but only slightly,  while the~TP relation meets only the~outer resonant point frequency interval.

The~RP--TP combination of the~oscillatory modes in the~RS~model implies parameters of the~neutron star that are quite acceptable in the~lower end of the~allowed ranges of spin and mass. Comparing our results with the~allowed ranges of spin (Eq.~\ref{limit-na-spin-z-rotace}) and mass (Eq.~\ref{limit-na-M-z-rotace}) we can see that the~RP--TP combination could work for $0.3 \lesssim a \lesssim 0.4$ and mass $  2.25 \lesssim M/\mathrm{M}_{\odot} \lesssim 2.5$.

The~physical explanation of the~RP--TP model could be relatively very simple. A~hot spot is oscillating in both vertical and radial directions and its radiation is modified by the~frequencies $\nu_{\theta}$ and $\nu_{\theta} - \nu_r$; approaching the~3\,:\,2 resonant point, oscillations in the~vertical direction are successively damped due to non-linear (e.g., tidal) effects and the~radial oscillations are enforced. Then the~Keplerian and the~radial epicyclic frequencies become important.

\subsubsection{Combination of the~RP1 and TP1 models}

Finally, we discuss one important case of combinations of modifications of the~RP and TP frequency relations. We do not follow all the~details, presenting only the~results. The~TP1 model has to be related to the~inner resonant point, while the~RP1 model has to be related to the~outer resonant point. For the~frequency scatter at the~resonant points, given by the~Eqs.~(\ref{freu}) and (\ref{frei}), we deduce the~range of allowed values of the~mass and spin of the~neutron star given by
\begin{equation}
             0.10 < a < 0.34 \,,\quad
             1.94 < \frac{M}{\mathrm{M}_{\odot}} < 2.36 \,.
\end{equation}
The~central point of the~frequency ranges implies the~central estimates of the~neutron star parameters to be $a=0.23$ and  $M = 2.15\,\mathrm{M}_{\odot}$.

The~RP1--TP1 combination of the~oscillatory modes in the~RS~model implies complete ranges of parameters of the~neutron star that are in a~good accord with the~theoretical limits on the~spin (Eq.~\ref{limit-na-spin-z-rotace}) and mass (Eq.~\ref{limit-na-M-z-rotace}) of the~neutron star in 4U~1636-53. This combination could thus be considered as one of the~best candidates for explaining the~observed data of the~HF~QPOs.

We can see from Table~\ref{Tab-M-a-pro-RPaTPmodely} where all the~results are presented that some combinations of the~frequency relations are clearly excluded by the~theoretical limits on mass and spin of neutron stars, while other give quite reasonable restrictions on the~mass and spin parameters of the~neutron star present in the~atoll source 4U~1636$-$53. We plan to test further the~acceptable combinations of the~frequency relations by fitting the~relations to the~observational data related to the~vicinity of the~resonant points. Our preliminary results indicate that the~improvement of the~fitting procedure precision could be really relevant. We believe that than we are able to fix more precisely the~proper combination of the~frequency relations.

\section{Conclusions}

The~RS~model has been tested for the~atoll source 4U~1636$-$53 demonstrating possible two resonant points in the~observed data. For relevant pairs of the~oscillatory frequency relations the~range of allowed values of the~mass and dimensionless spin of the~neutron star are determined giving in some cases acceptable pairs of frequency relations, while some other pairs are excluded because of predicting unacceptable values of the~spin and/or mass of the~neutron star at 4U~1636$-$53. We focused our attention on the~test of the~frequency relations containing geodetical orbital and epicyclic frequencies or some combinations of these frequencies. It should be noted that the~cause of the~switch of the~pairs of the~oscillatory modes is not necessarily tied to the~resonant phenomena related to the~oscillations governed by the~frequencies of the~geodetical motion, as the~switch can be related, e.g., to the~influence of the~magnetic field of the~neutron star, or the~radiation coming from the~surface of the~neutron star. Moreover, in some sources, e.g., the~Circinus X-1, we cannot exclude the~possibility of chaotic changes of frequency relation pairs in a~given fixed interval of observed frequencies. We study the~resonant phenomena first, leaving other cases to the~future studies.

Generally, the~RP--TP combination of the~RS~model, and its modifications, enable acceptable explanation of the~observational data for 4U~1636$-$53 source. This should be, however, explicitly tested by fitting procedure applied to the~observed twin HF~QPO sequences related to the~resonant points. We plan to make such a~test in a~future work and also to test the~RS~model in the~case of some other atoll (4U~1608$-$52) or Z (Circinus~X-1) sources, containing a~neutron (quark) star, with observational data indicating possible existence of two resonant points, and to estimate allowed values of the~spin and mass of the~neutron (quark) stars.

In the~special situations related to accreting neutron stars with near-maximum masses the~Kerr metric can be well applied in calculating the~orbital and epicyclic geodetical frequencies, as has been done in the~present paper, where the~results of the~mass and spin interval findings are in agreement with the~assumption of near-maximum masses of the~neutron stars. In general situations, when the~neutron star mass is not close to its maximum value allowed by the~equation of state, or when a precise fitting procedure of the~observational data is needed, the~Hartle--Thorne geometry describing rotating neutron stars has to be considered and the~orbital and epicyclic frequencies reflecting influence of mass, spin and quadrupole moment of the~neutron star has to be used. Nevertheless, the~role of the~quadrupole moment is relevant only very close to the~inner edge of the~accretion disk \citep{Tor-etal:2010:ASTRJ2:MassConstraints}.

\sloppy
\vspace{3ex}
\noindent
{\small{\bf{Acknowledgements.}}
We would like to express our gratitude to the~Czech grants GA\v{C}R~202/09/0772, GA\v{C}R~205/09/H033 and the~internal grants of the~Silesian University Opava FPF SGS/1,2/2010. The~authors further acknowledge the~pro\-ject Supporting Integration with the~International Theoretical and Observational Research Network in Relativistic Astrophysics of Compact Objects, reg. no. CZ.{\-}1.07/2.3.00/20.0071, supported by Operational Programme \emph{Education for Competitiveness} funded by Structural Funds of the~European Union and state budget of the~Czech Republic.}


\providecommand{\uv}[1]{\glqq#1\grqq}

\end{document}